# Observation of Diode Behavior and Gate Voltage Control of Hybrid Plasmon-Phonon Polaritons in Graphene-Hexagonal Boron Nitride Heterostructures

Francisco C. B. Maia,*,[1] Brian T. O'Callahan,[2] Alisson R. Cadore,[3] Ingrid D. Barcelos,[1,3] Leonardo C. Campos,[3] Kenji Watanabe,[4] Takashi Taniguchi,[4] Christoph Deneke,[5,6] Alexey Belyanin,[7] Markus B. Raschke,[2] Raul O. Freitas*,[1]

[1]Brazilian Synchrotron Light Laboratory (LNLS), Brazilian Center for Research in Energy and Materials (CNPEM), Zip Code 13083-970, Campinas, Sao Paulo, Brazil.

[2]Department of Physics, Department of Chemistry, and Joint Institute for Lab Astrophysics (JILA), University of Colorado, Boulder, Colorado 80309, United States.

[3]Department of Physics, Federal University of Minas Gerais, 30123-970 – Belo Horizonte, Minas Gerais, Brazil.

[4]Advanced Materials Laboratory, National Institute for Materials Science, 305-0044 – Namiki, Tsukuba, Japan.

[5] Brazilian Nanotechnology National Laboratory (LNNano), Brazilian Center for Research in Energy and Materials (CNPEM), Zip Code 13083-970, Campinas, Sao Paulo, Brazil

[6] Applied Physics Department, Gleb Wataghin Physics Institute, University of Campinas (Unicamp), Zip Code 13083-859, Campinas, SP, Brasil.

[7]Department of Physics & Astronomy, Texas A&M University, College Station, Texas 77843-4242, United States.

*Corresponding authors

Light-matter interaction in two-dimension photonic materials[1–5] allows for confinement and control of free-space radiation on sub-wavelength scales[6]. Most notably, the van der Waals heterostructure obtained by stacking graphene (G) and hexagonal Boron Nitride (hBN) can provide for hybrid[2] hyperbolic plasmon phonon-polaritons (HP$^3$) [7,8]. Here, we present a polariton diode effect and low-bias control of HP$^3$ modes confined in G-hBN. Using broadband infrared synchrotron radiation coupled to a scattering-type near-field optical microscope[9–11], we launch HP$^3$ waves over both hBN Reststrahlen bands[12] and observe the unidirectional propagation of HP$^3$ modes at in-plane heterointerfaces associated with the transition between different substrate dielectrics. By electric gating we further control the HP$^3$ hybridization modifying the coupling between the continuum graphene plasmons and the discrete hyperbolic phonon polaritons of hBN as described by an extended Fano model[13–15]. This is the first demonstration of unidirectional control of polariton propagation, with break in reflection/transmission symmetry for HP$^3$ modes. G-hBN and related hyperbolic metamaterial nanostructures can therefore provide the basis for novel logic devices of on-chip nano-optics communication and computing.

Photonics based on two dimension (2D) materials[3–5] exhibits a plethora of optical nanoscale phenomena including the ability to confine free-space radiation to the deep sub-wavelength scale[6]. Remarkably, photonic crystals and metamaterials built from hybrid 2D heterostructures provide for qualitatively new ways of light-matter interaction on the nanoscale[1,2] and are approaching their ultimate purpose as components for novel nano-photonic devices[4,16]. Nano-electronics has recently reached a pivotal stage with the realization of atomically thin p-n junctions[17]. Hence, one can readily envisage 2D photonic structures as logic elements for future nano-optical circuits.

Graphene has excelled at converting light into surface plasmon polaritons (SPPs)[18,19] with deep-subdiffraction wavelength gate tunability[20,21]. Similarly, mid-infrared free-space radiation can couple to subwavelength-confined hyperbolic phonon polaritons (HPhPs) in hBN[8,22] that are able to propagate up to 20 times the typical transmission distances of graphene plasmons[22]. Merging the optical attributes of graphene and hBN, the G-hBN heterostructure constitutes an electromagnetic hybrid[2] with enhanced opto-electronic performance compared to its standalone constituents. In G-hBN, coupling between graphene SPPs and HPhPs of hBN creates the hybrid hyperbolic plasmon phonon-polaritons ($HP^3$) modes[7,8] with gate-tunable wavelengths[7]. G-hBN not only supports graphene plasmons with reduced damping[23] but also $HP^3$s with propagation lengths up to 2 times greater than HPhPs in bare hBN[7]. Substrate influence on the propagating $HP^3$ modes was theoretically suggested[24]. To date, however, a G-hBN-based waveguide, with diode-like functionality and full electro-optical control of light confinement, is still missing.

In this work, we report optical phonon-polariton confinement, a diode effect driven by asymmetric power flow of $HP^3$ modes, and low-voltage bias-control of $HP^3$ modes in G-hBN nano-device mounted on a air-Au metasurface (Fig. 1a). This studied is performed by spatio-spectral nano-imaging from scattering scanning near-field microscopy (s-SNOM) using synchrotron infrared nano-spectroscopy (SINS)[9–11].

Rectification of phonons, often measured through thermal transport, has been predicted in 1D chain of atoms with a mass gradient[25] and across interfaces between materials with differing mass[26], and has been reported in boron nitride nanotubes[27]. Asymmetry of reflection coefficients of graphene SPPs has been observed across grain boundaries due to a variation of the doping density and corresponding

SPP momentum[28]. By configuring the G-hBN/Au heterostructure as a gate-controlled device (Fig. 1a), we externally modulate the graphene plasmons and achieve fine tuning of the optical response of the type I HP$^3$s (Fig. 2). We see reflection of the HP$^3$ modes in the hBN crystal at the transitions between air to Au substrates (Fig. 3-4). Such reflection is caused by a mismatch between the HP$^3$ wavevectors and a corresponding change in mode volume for both type I and type II bands. In fact, air and Au modify the dielectric environment of the hBN crystal. As a result, G-hBN/Au and G-hBN/air form a diode-like junction for polaritons that, asymmetrically, regulates the power flow of the HP$^3$ modes.

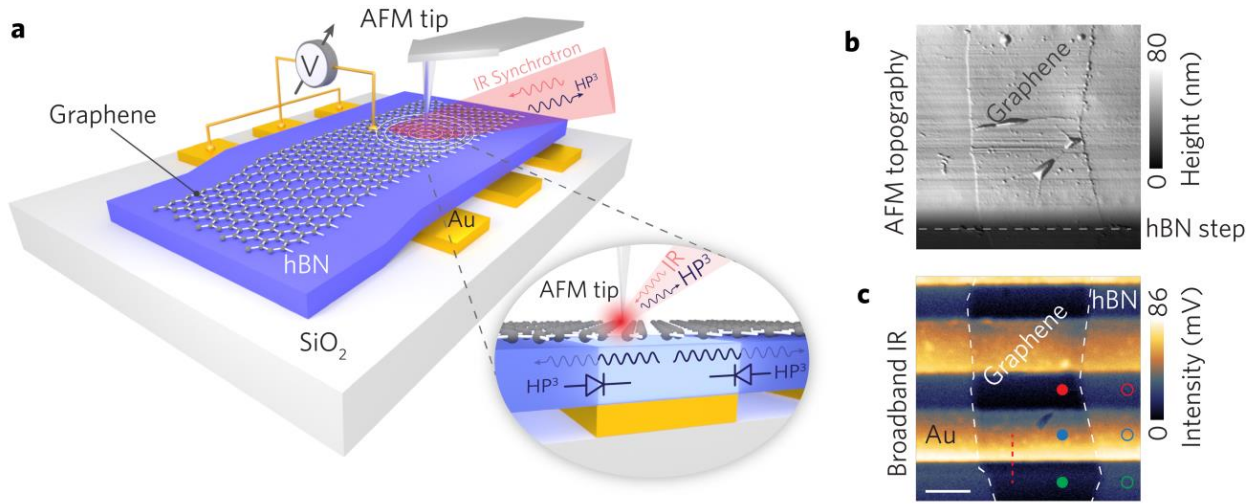

**Figure 1| G-hBN device structure and experiment configuration. a,** Schematic of the G-hBN heterostructure device with different dielectric media (SiO$_2$, Au and air) and electric gating setup. Using SINS, HP$^3$ modes are launched and scattered-detected by the metallic AFM tip. Inset: illustration of the asymmetric polariton propagation for the HP$^3$ modes. **b,** AFM topography of the hBN/SiO$_2$ edge region, the hBN on Au pads area, and the graphene. **c,** Simultaneously acquired broadband infrared near-field image, with dielectric contrast of Au contacts under the 20 nm of hBN, and the graphene sheet (scale bar represents 1 μm). Circles indicate location of near-field spectra displayed in Fig. 2. Red dashed line locates the spectral linescan shown in Fig. 3b.

The G-hBN heterostructure is assembled onto SiO$_2$, Au stripes, and across air gaps by standard mechanical exfoliation using adhesive tape (for details see methods and Refs. [29,30]). The flatness and quality of the hBN and graphene surfaces were characterized through a combination of optical microscopy, atomic force microscopy (AFM), and Raman spectroscopy. An electrode on top of the graphene region allowed for charge injection by tuning the gate voltage as illustrated (Fig. 1a). The graphene is identified by the AFM topography (Fig. 1b) and in the spectrally integrated broadband s-SNOM image (Fig. 1c), with corresponding dielectric contrast from the underneath Au contacts. The combined AFM/s-SNOM imaging confirms that the 20 nm hBN crystal smoothly bridges the Au contacts, thus forming free-standing G-hBN/air regions.

SINS, carried out at the Brazilian Synchrotron Light Laboratory (LNLS), provided for tip-based nanolocalized spectroscopy, imaging and spatio-spectral probing as established for graphene SPPs[20,21,28], hBN HPhPs[22,31] and G-hBN HP³s[7,8]. In its implementation with synchrotron radiation, SINS operates with high spectral irradiance through the mid-IR range (750 cm$^{-1}$ to 3000 cm$^{-1}$)[11] and it is capable of probing both hBN Reststrahlen bands simultaneously.

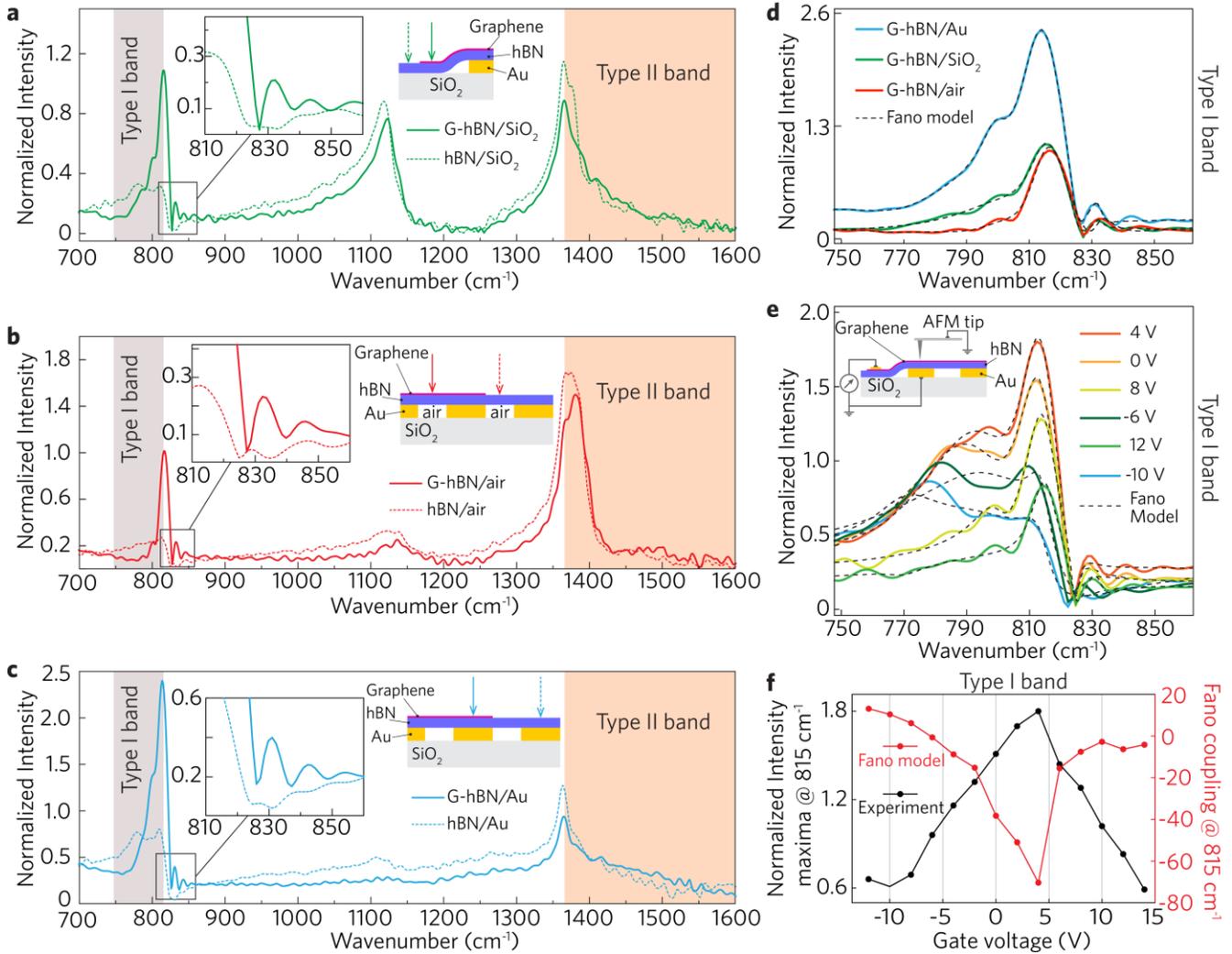

**Figure 2| Substrate and gate modulation of the G-hBN near-field response.** SINS point spectra of G-hBN and hBN lying on SiO$_2$ (**a**), air (**b**, freestanding structure) and Au (**c**), whose measurement locations are indicated by the inset cartoons in **a**-**c**. The type I (740 to 820 cm$^{-1}$) and type II (1350 to 1610 cm$^{-1}$) bands are highlighted in **a**-**c**. Spectra insets in **a**-**c**: zoom into spectral

oscillations near the edge of the type I band associated with the graphene/hBN interaction. **d**, Type I band optical response of the G-hBN heterostructure lying on Au, SiO$_2$ and air substrates with characteristic Fano lineshapes and corresponding numerical fits (dashed black curves). **e**, Gate tuning of the type I band optical response and device schematic for controlling voltage bias between graphene and Au (inset). **f**, Normalized intensity maxima (from curves in **e**) and Fano coupling values of the 815 cm$^{-1}$ mode as a function of the applied gate voltages.

Figs. 2a-c show SINS point spectra acquired at the device locations indicated in the respective insets. The spectra in Fig. 2a, taken on the hBN/SiO$_2$ structure with and without graphene, are characterized by three resonances assigned to the hBN type I band (650 – 820 cm$^{-1}$), SiO$_2$ surface phonon-polariton near 1130 cm$^{-1}$, and hBN type II band (1365 – 1610 cm$^{-1}$). The type I band presents increase in intensity for the G-hBN structure due to the HP$^3$ coupling as seen previously[8]. In the Fig. 2b for air substrate, the type II band is dominant. Yet, on G-hBN/air the type I band is more intense than in hBN/air. Fig. 2c presents spectra with the Au substrate revealing an expressive increase of the type I band for the G-hBN/Au compared to hBN/Au. Zoomed spectral insets in Fig. 2a-c show oscillations, on the higher wavenumber side of the type I band, specifically associated with graphene presence.

Concerning the type II band, the comparison of hBN/SiO$_2$ and G-hBN/SiO$_2$ spectra (Fig. 2a) reveals that graphene prompts a slight blue shift to the transverse optical (TO) mode frequency at 1365 cm$^{-1}$. The hBN/air and G-hBN/air (Fig. 2b), however, exhibit increased intensities with respect to both SiO$_2$ (Fig. 2a) and Au (Fig. 2c). As shown in the Figure 4c and in ref. 32, type II HP$^3$ modes present higher momenta (higher confinement) on Au than on air and SiO$_2$ substrates. This explains the reduced intensity of the type II band modes on Au substrate since it is expected less efficiency for launching highly confined HP$^3$ modes[33]. In contrast, the type I band changes are substantially distinct. Fig. 2a demonstrates a HP$^3$ coupling increase to the type I band of more than 200% for G-hBN/SiO$_2$ with respect to hBN/SiO$_2$, whilst, previous reports[7,8] show relative increase below 50%. The HP$^3$ amplitude increase also manifests itself for G-hBN/air with respect to hBN/air (Fig. 2b). In Fig. 2c, G-hBN/Au shows amplitude modes with distinct line-shape and almost a factor of three times higher intensity than hBN/Au. Comparatively, G-hBN/Au yields the largest amplitude increase, indicating that Au contributes to the HP$^3$ coupling (see Supplementary Information). The spectral oscillations (spectral insets of Fig. 2a-c), outside of the type I band and seen only for the G-hBN heterostructure, are a manifestation of hybridized modes. The interaction of broadband SPPs of graphene with HPhPs of hBN, resulting in amplitude changes and line-shape asymmetry, are described here by the Fano model[13–15].

The analytic expression of the Fano resonance [34] $I_{Fano}(\omega)$ is given by

$$I_{Fano}(\omega) \propto \frac{p(f \cdot \Gamma + (\omega - \omega_0))^2}{(\omega - \omega_0)^2 + \Gamma^2}, \qquad \text{eq. 1}$$

with excitation frequency $\omega$, phonon frequency $\omega_0$, damping $\Gamma$, transition dipole strength of coupled mode $p$, and dimensionless Fano coupling $f$, which corresponds to the dipole strength of the transition between the discrete and continuum states. Thus, $f$ provides a measure for the HP$^3$ mode coupling. We extend Eq. 1 to the case of interaction of multiple HPhPs modes (index j) with the graphene quasi-continuum SPPs, and a non-resonant background $A_{nr}$ (see Supplementary Information), resulting in the expression:

$$I_{HP^3}(\omega) \propto \left| \sum_j \frac{\sqrt{p_j}\left(f_j \Gamma_j + (\omega - \omega_{0j})\right)}{(\omega - \omega_{0j}) + i\Gamma_j} + A_{nr} \right|^2 \qquad \text{eq. 2}$$

We use eq. 2 to fit, with good agreement, the type I band of the G-hBN spectra on the different substrates (Fig. 2d), and the G-hBN/Au spectra for different gate voltages (Fig. 2e). For all fits the number of modes and their respective resonance frequencies are kept fixed with three modes inside the type I band at 770, 800 and 815 cm$^{-1}$ and two outside this band at 824 and 832 cm$^{-1}$.

The substrate mostly induces changes in $f$ (see the Supplementary Information). A large negative Fano coupling is found for the mode at 815 cm$^{-1}$ of the G-hBN/Au spectrum indicating pronounced coupling of the primarily out-of-plane phonons with the graphene and Au plasmons. G-hBN/SiO$_2$ and G-hBN/air present $f$ of equal magnitude for the same mode, yet, smaller than for G-hBN/Au. This suggests that the coupling is much weaker for SiO$_2$ and air substrates than for the Au substrate.

By gating G-hBN/Au as represented in the Fig. 2e schematic inset, we vary the electric field crossing the hBN crystal. The gate-driven electric field, which is parallel to the out-of-plane component of the hBN dielectric function, tunes the carrier density in graphene yielding the systematic modulations of the amplitude of the type I band (out-of-plane modes) as seen in the Fig. 2e. Since this electric field structure is orthogonal to the in-plane modes, negligible electric modulation is induced to the type II band. From -12 to 0 V an overall intensity increase occurs. From 0 to 4 V, despite a continuous increase of the 815 cm$^{-1}$ mode amplitude, an onset of amplitude attenuation occurs for the modes at 770 and 800 cm$^{-1}$. From 4 to 12 V, the 815 cm$^{-1}$ mode amplitude decreases almost monotonically, while the other two modes gradually quench. The corresponding change in the measured intensity of the 815 cm$^{-1}$ mode as a function of the gate voltage is shown in the Fig. 2f, with the maximum at 4V. Such profile is analogous to characteristic curves of resistance versus gate voltage for graphene that permits identifying the Dirac point, the intrinsic doping level and assessing the mobility of charge carriers. Thus, the maximum at 4V, which implies intrinsic p-type doping that is likely to result from polymer resist residues, unveils that the HP$^3$ coupling is favored at the Dirac point where the number of electrons and holes are equal, i.e. the charge neutrality point. And, when the bias induces an unbalance in the number of electrons and holes, the mode amplitude decreases.

From the fittings in Fig. 2e, we obtain the Fano coupling dependence for the 815 cm$^{-1}$ mode as a function of gate voltage as plotted in Fig. 2f (red circles). At 4V, *f* has the largest negative value signifying that, in the HP$^3$ coupling, the phonon contribution dominates over the electronic one [14]. However, as the voltage is tuned away from the charge neutrality point (4V), *f* tends to smaller values representing stronger contributions from free electrons in graphene. Therefore, the combination of experimental amplitude/Fano factors of the 815 cm$^{-1}$ mode demonstrates that the gate tunability allows for modifications in the HP$^3$ coupling by the electrostatic-driven changes in graphene plasmons. We also note that the asymmetry of the *f* values around 4 V (Fig. 2f) indicates that electrons and holes have dissimilar influence on the HP$^3$ coupling.

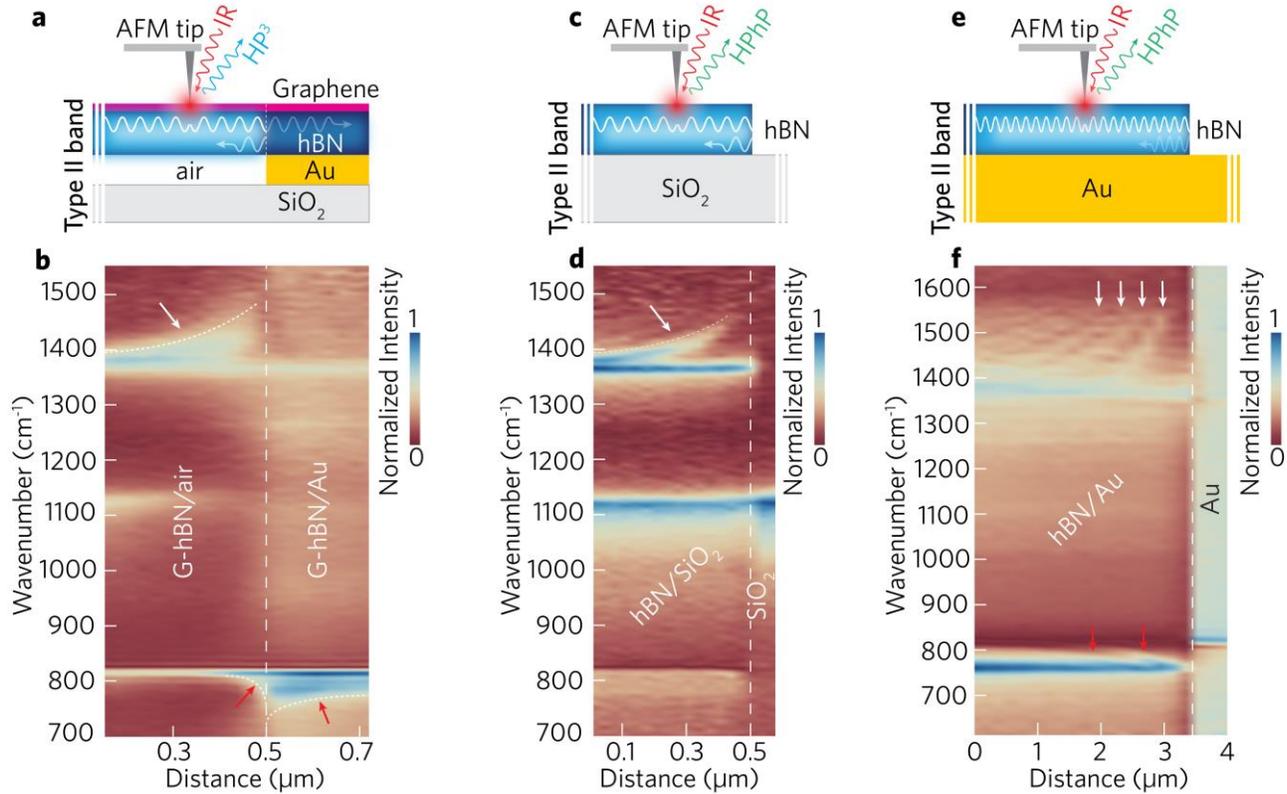

**Figure 3| HPhP and HP³ modes reflecting at the hBN crystal edge and at the air-Au dielectric interface**. **a**, **c** and **e**, schemes of polaritonic waves propagation, respectively, for the G-hBN heterostructure on the air-Au substrate transition, 20 nm thick hBN crystal on SiO₂ and 240 nm thick hBN crystal on Au film. **b**, spectral linescans across the G-hBN/air to G-hBN/Au regions, for zero gate voltage, (scheme **a**) showing regions of intensity depletion/increase for the type I (on Au side) and type II (on air side) as indicated by arrows. The SiO₂ surface phonon is seen at ~1130 cm⁻¹. **d**, **f**, Spectral linescans, respectively, across hBN/SiO₂ edge (scheme **b**) and the hBN/Au edge (scheme **c**).

Fig. 3 compares, in the form of spectral linescans, HPhP modes reflected at the edge of the hBN crystal on SiO$_2$ and Au substrates (Fig. 3d,f) with the propagation of HP$^3$ modes in the G-hBN heterostructure lying on the air-Au metasurface (Fig. 3b). Whereas the spectral linescan of hBN/SiO$_2$ shows only modest features for the type I band, the type II band present the typical spatio-spectral branches (arrows in the Fig. 3d) corresponding to the HPhP reflection from the crystal edges[22,31]. A spectral linescan across the edge of a 240 nm thick hBN crystal on Au film is presented in the Fig. 3f. The dense pattern of fringes (white arrows in the Fig. 3f) in the type II band of the hBN/Au demonstrates that HPhP modes are excited by the tip on Au substrate, propagate to the crystal edge, reflect and form standing waves as in the hBN/SiO$_2$ case (Fig. 3d). Despite being a minor effect, HPhPs propagation/reflection can also be inferred in the type I band of hBN/Au (2 red arrows, Fig. 3f. See Supplementary Information). Analogous effects rise in the spectral linescan of the G-hBN heterostructure bridging the air-Au metasurface (Fig. 3b). The type II HP$^3$ modes, within 200 nm from the air-Au substrate transition, feature on the air side an intensity increase/depletion similar to the one observed in Fig. 3d and 3f related to edge-reflected type II modes. Yet, the HP$^3$ modes of type I present intensity increase/depletion from the vicinity of the air-Au transition (red arrows), on the Au side (Fig. 3b). As mentioned, to date, such patterns in spectral linescans have been exclusively ascribed to standing waves formed by tip-launched circular waves that propagate, reflect and interfere with incoming ones[22]. Hence, the analogy of spectral linescans of G-hBN on the air-Au metasurface (Fig. 3b), hBN/SiO$_2$ (Fig. 3d) and hBN/Au (Fig. 3f) permit attributing the intensity increase/depletion near the air-Au junction to reflection of HP$^3$ modes. Moreover, such effect can be modulated via gating (see Supplementary Information).

Hence, from a qualitative understanding, reflection of the in-plane polarized modes (type II modes) occurs when approaching the air-Au substrate transition from and air side (Fig. 3a). On Au side, the reflection of this band is minorized because of the higher confinement of the type II HP$^3$ modes on this substrate and unfavorable mode coupling at the metasurface, as explained by the reflection coefficient model in the Fig. 4d. The opposite trend is noted in the type I band: the reflection of the out-of-plane modes occurs, prominently in the Au region and very moderately on the air side.

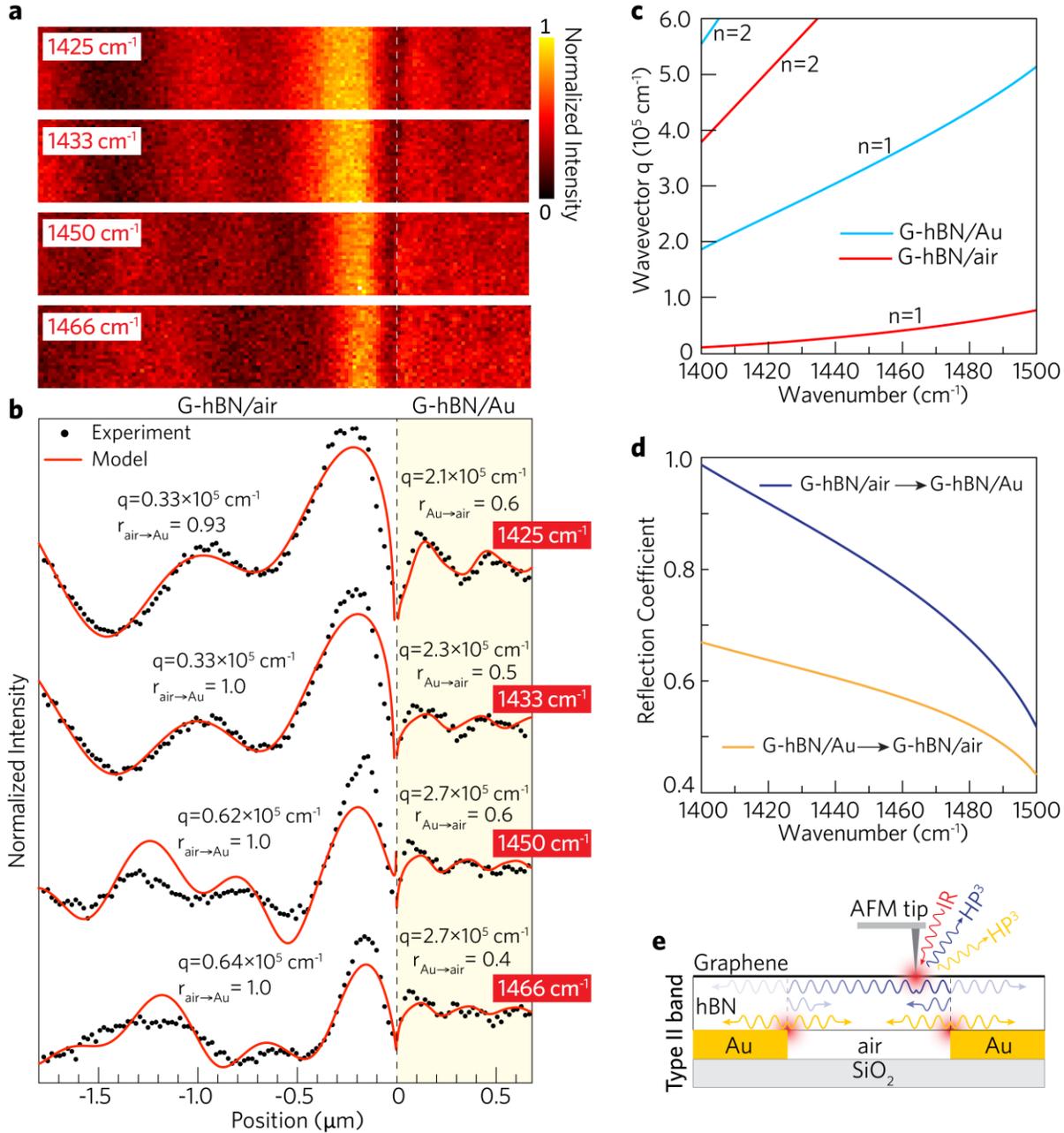

**Figure 4| Narrow-band intensity maps, intensity profiles, dispersion and reflection coefficients for the type II HP³modes propagating over the Au-air junction. a,** Intensity maps of a 35 nm thick G-hBN, on a 0.5 µm x 2.5 µm region with the underneath air-Au metasurface (air-Au transition is marked by the white dashed lines), for different excitation wavenumbers in the type II band and zero gate voltage. **b,** Intensity profiles extracted from the narrow-band intensity maps in **a**. Dots are experimental data and red lines are fits from the model considering the tip and the Au edges as launchers of HP³ waves as shown in the cartoon in **e**. **c,** Frequency-momentum dispersion relation of the type II HP³ modes regarding the Au and air substrate media and measured 160 meV Fermi level for graphene. **d,** Reflection coefficient calculated from eq. 3 for type II HP³ modes propagating from G-hBN/air to G-hBN/Au regions (blue curve) and in the opposite direction (yellow curve). **e,** Scheme of the HP³ waves generated by the tip and Au edges antennas. These waves are considered in the model for fitting the oscillations in **a**. In this model: tip launches isotropically circular waves that travel to and reflect at the G-hBN/Air and G-hBN/Au left and right junctions. Plane waves are excited by Au edges. The resulting optical field $\xi_{Opt}$ originates from the interference of those waves in **e**.

The propagation of the HP$^3$ modes on the air-Au metasurface can be understood by the power flow of their associated waves, which is given by the Poynting vector. The different mode structure in the G-hBN/air and the G-hBN/Au regions of the waveguide (G-hBN) imposes an asymmetry for the power flow of the HP$^3$ modes across the junction between the two regions. This mechanism represents a diode behavior for the HP$^3$ modes. The origin of the asymmetric power flow is confirmed by the calculated dispersion of polariton modes for each heterostructure (Fig. 4c). Note that the asymetry of the multimode power flow due to the mismatch of the polariton mode structure does not constitute Lorenz nonreciprocity, i.e., asymmetry in the scattering matrix between each pair of modes. There is no reason for the violation of Lorenz reciprocity since our system is linear, time-independent, and has no magnetic field applied. [35]

We calculate the frequency-momentum dispersion relation for the type II HP$^3$s for Au and air substrates (Fig. 4c) through the reflection coefficient of the surface of the air/G-hBN/substrate layered system. Polariton resonances occur at the divergences of this reflection coefficient [22,36]. Due to its hyperbolic dispersion, hBN supports multiple discrete polariton dispersion branches with momenta $q_n$ indexed by integers $n$ = 1,2,… so that $q_n$ increases with increasing $n$ [7,37,38]. The dispersion relations of the first two branches for each substrate are plotted in Fig. 4c for the type II band (for type I band, see Supplementary Information). For the type II band, the Au substrate promotes higher values of in-plane HP$^3$ momenta than the air region for the same mode index $n$. Conversely, type I modes on the air region have higher in-plane HP$^3$ momenta than on the Au region (Supplementary Information). Hence, the structure of polariton modes in G-hBN/air is different from that in G-hBN/Au leading to a mismatch of momenta at the junction. Since the dispersion in bulk hBN is hyperbolic, any polariton with a high in-plane momentum has also a high vertical component of momentum. For a waveguide mode, this translates into a high degree of confinement (ratio between free-space wavelength to polariton wavelength in the crystal). It is then straightforward to see that a given transverse mode from a high-momentum (strong-confinement) region scatters with high efficiency into a set of transverse modes in a region with a lower confinement, which is the case of the in-plane polarized HP$^3$ modes propagating from Au to air sides. Whereas, the same process originated from the opposite side is much less efficient (see also ref. [35] ).

To compute the expected polariton reflection properties, we consider in-plane polarized polariton modes travelling from a region with in-plane polariton momenta $q_n$ into a region with different momenta $q'_n$. Since the tip primarily excites the n = 1 mode, we only consider the transmission of the mode with momentum $q_1$ across the substrate interface. In this case, the polariton reflection coefficient can be approximately calculated as

$$r = \frac{q'_1-q_1}{q'_1+q_1} \prod_{n=2}^{s} \frac{(1-q_1/q'_n)(1+q_1/q_n)}{(1-q_1/q_n)(1+q_1/q'_n)} \quad \text{eq. 3}$$

where the product is over mode indices up to integer s [39]. To simplify the calculation, we neglect contributions from mode indices with $n > 2$, since the first two modes with the smallest momenta will dominate the reflection coefficient. The results are shown in Fig. 4d. The calculated reflection coefficients show pronounced asymmetry, with higher values for polaritons travelling from a low-momentum region to a high-momentum region. This is because polaritons with low $q_1$ can only efficiently couple to the lowest polariton branch with momentum $q'_1$. However, for the case where $q'_1 < q_1 < q'_2$, there is significant transmission into both polariton branches. While the excitation of the n = 2 mode has been observed previously[40], its excitation efficiency and, therefore, its role in the diode behavior is weak compared to the n = 1 mode. Thus, there is better coupling efficiency for polaritons travelling from a high momentum region to a low momentum region. This steers to strong reflection of type II modes when travelling from the air region to the hBN/Au region, and conversely, of type I modes travelling from the Au to the air substrates.

The diode prediction is further attested by a quantitative analysis of the narrow band intensity maps (Fig. 4a) extracted from the narrow-band intensity maps, obtained from a hyperspectral image (see details in Methods), for the type II band (for type I band, see Supplementary Information) of a region with the G-hBN/Au and G-hBN/air junction. Fig 4a reveals, for 4 excitation frequencies $\omega$, different patterns of maxima and minima in the air and Au regions, which produce the oscillating intensity profiles (see details in Methods) in the Fig. 4b. Such oscillations have a non-trivial behavior that cannot be straightforwardly ascribed to pure plane waves launched by Au edges either to pure circular waves launched by the tip[32,41,42]. In our case, those oscillations correspond to the resulting optical field $\xi_{Opt}$ created in the G-hBN heterostructure by the interference of HP$^3$ circular waves from the tip ($\xi_{Tip}$, eq 4.) and plane waves from the Au edges ($\xi_{Au\ edges}$, eq. 5). Thus, $|\xi_{Opt}|^2 = ||\xi_{Tip} + \xi_{Au\ edges}||^2$ is the scattered optical field that reaches the detector. In eq. 4, $A$, $\alpha$ and $r$ are, respectively, amplitude, phase and the reflection coefficient (see supplement for more details). In eq. 5 $B$ and $\beta$ are, respectively, amplitude and phase of the Au edge plane waves. The term $Ce^{-i\eta}$ is the contribution of a non-resonant background originating from other scattering centers coexisting in the far-field illuminated confocal spot. All waves in eq. 4 and 5 have the same complex momentum expressed by the real $q$ and imaginary $\kappa$ parts.

$$\xi_{Tip} = A\frac{e^{-i\alpha}}{\sqrt{x}}\left(e^{i(q+i\kappa)x} - r \times e^{-i(q-i\kappa)2x} + e^{-i(q-i\kappa)x} - r \times e^{i(q+i\kappa)2(x_{edge}-x)}\right) \text{ eq. 4}$$

$$\xi_{Au\ edges} = Be^{-i\beta}\left(e^{-i(q-i\kappa)x} - e^{i(q+i\kappa)2(x_{edge}-x)}\right) + Ce^{-i\eta} \text{ eq. 5}$$

We fit the oscillating intensity profiles in the Fig. 4b to model given by the $|\xi_{Opt}|^2$ expression. The resulting fittings (Fig. 4b) reasonably agree with the experimental data. Using the mode-extracted amplitude of each wave, we calculate, for each $\omega$, the launcher efficiency[41] for the tip $\sigma_{Tip} = 2\pi[2A(1+r)]^2$ and the Au edges $\sigma_{Edge} = B^2$. As shown in the Table 1, the tip is the dominant launcher regardless the substrate ($\sigma_{Tip} > \sigma_{Edge}$). Model-extractel $q$ values (indicated in Fig. 4b next to respective intensity profiles) reasonably match the calculated dispersion relations (Fig. 4c) for both substrates, confirming that the HP$^3$ modes for type II band are more confined (higher momenta values) in Au than in air regions[32]. Importantly, the model-extract reflection coefficients agree with the theoretical predictions (Fig. 4d) for all $\omega$ and for both substrates. One can observe the asymmetric reflection coefficient values, for example: at $\omega$ = 1425 cm$^{-1}$, $r_{air \to Au} = 0.93$, when the tip is on the air side, signifies that most of power remains in the air substrate side. Whilst, when the tip is on the Au side, $r_{Au \to air} = 0.6$ indicates that substantial part of the power transmits to the air side. Note that such reflection and changes to the momenta and damping (see Supplementary Information) are only caused by the Au pad beneath the 35 nm thick G-hBN without any physical discontinuity induced to the crystal. Thereby, G-hBN/Au can be seen as an electronic barrier[43] wherein the associated plasmonic activities of Au and graphene modify the dielectric environment compared to the G-hBN/air. Furthermore, the convergence between theory and the model-extracted values confirms an unidirectional regulation of the HP$^3$ power flow mediated by the G-hBN/air and G-hBN/Au junction, which configures the diode behavior.

We note again that the observed diode effect, i.e., unidirectional regulation of the power flow, differs from optical isolation, as implemented in Faraday isolators, since our experimental conditions do not break Lorenz reciprocity. This diode effect is a consequence of the local excitation of HP$^3$s with mode index $n = 1$, which can couple to higher order modes during transmission to regions of different polariton confinement. It is this coupling which possesses asymmetry. However, the process is still reversible in the following sense when the $n = 1$ tip-excited polaritons in region A experience the substrate mediated dielectric junctions, while, propagating towards region B, they transmit into a superposition of polariton modes with mode indices $n \geq 1$ propagating in region B. If that same superposition with the same relative phase relation is incident from region B on the boundary from the opposite direction, they will combine to create a pure $n = 1$ transmitted mode propagating in region A.

The gating control demonstrates the nature of the HP$^3$ coupling between quasi-continuum SPPs of graphene interacting with discrete HPhPs of hBN. In particular, the observed tunability reveals the possibility of a nanoscale opto-electronic device with logic functionality in analogy to electronic transistors and diodes. Furthermore, the control of the HP$^3$ mode at 815 cm$^{-1}$ of the type I band provides fine tuning of a high polarized HP$^3$ mode. This, combined with the diode effect, creates the possibility of an externally-driven, polarization sensitive device for subwavelength light. In summary, we present and quantitatively explain substrate-mediated unidirectional power flow and gating control of HP$^3$ modes in G-hBN heterostructure. This demonstration of a 2D polaritonic device with diode-like properties and with accessible control of light confinement marks a prime step towards the realization of nanoscale on-chip polaritonic elements for optical communication and computing applications. In general, our experimental findings can be adapted for other 2D polar materials and alternative optical junctions can be designed by simply varying the in-plane form-factor of the heterostructure substrate. Our results open a route for the realization of more complex 2D photonic devices such as detectors, modulators and metamaterial systems in the mid-infrared wavelength range.

# Methods

The Au stripes (60 nm thick) were designed by standard electron-beam lithography and thermal metal deposition. In sequence, single layer of mechanically exfoliated graphene was transferred atop of hBN flake forming the G-hBN/SiO$_2$, G-hBN/Au and G-hBN/air heterostructures constructing the sample structure shown in Fig 1a. The graphene partially covered the flake leaving areas with bare hBN. After each material transference, the sample was submitted to thermal annealing at 350 °C with constant flow of Ar:H$_2$ (300:700sccm) for 3.5 h to remove organic residues. The heterostructures measured in the Fig. 2 and 3a corresponded to a 20 nm thich hBN, while, in the Fig. 4, to a 35 nm thick crystal.

In the IR beamline of the LNLS, the infrared radiation extracted from a 1.67 T bending magnet is collimated (described in the ref. [44]) and then coupled to a commercial s-SNOM (NeaSnom, Neaspec GmbH). The IR synchrotron beam is focused on the metallic AFM tip shaft (NCPt-Arrow and PtSi-NCH, Nanoworld AG) which works as an extended antenna of about 25 nm apex radius. This antenna, which has a quasi-achromatic response in the mid-IR, is now an evanescent broadband source with dimensions comparable to the tip size. Operating in semi-contact mode AFM (tip frequencies from 250 to 350 kHz), the sample is brought in close proximity to the tip. The optical fields confined in the gap between tip apex and the sample gap generate an mutual polarization with amplitude and phase carrying the local sample dielectric response[45]. A Mercury Cadmium Telluride detector (MCT KLD-0.1, Kolmar Technologies Inc.) is responsible for the detection of the scattered fields from the tip-sample interaction volume. For background suppression, we perform lock-in detection of the MCT signal on the 2$^{nd}$ and 3$^{rd}$ harmonics of the tip frequency[46]. Phase-sensitive nano-spectroscopy is obtained by mounting the tip-sample stage in an asymmetric interferometric scheme which is also used to demultiplex the spectral response of the broadband beam in similar fashion as Fourier Transform IR spectroscopy (FTIR)[47,48]. We used 5 cm$^{-1}$ spectral resolution by setting 2 mm optical path difference (OPD) of travel for the scanning mirror. The spectra acquisition was done by integrating over 2048 points with 20.1 ms integration time per point. All spectra are normalized by a reference spectrum acquired from a film of gold with 100 nm sputtered on a silicon substrate. The images shown in Fig. 1 b and c were taken with 250 pixels ×250 pixels with 15 ms integration time per pixel.

Narrow-band maps shown in Fig. 4a were extracted from a full hyperspectral map (2D s-SNOM map where every tip position contains a broadband IR spectrum) of a 0.5 μm x 2.5 μm area on G-hBN over the air-Au substrate transition. This area was segmented in 25 pixels x 125 pixels. Each spectrum in the hyperspectral map resulted from 5 averages over the Fourier Transform of a interferogram acquired with 600 μm optical path difference length, yielding 16.6 cm$^{-1}$ spectral resolution, divided in 400 points with 5.7 ms integration time per point. The narrow-band intensity maps (Fig. 4a) were obtained by extracting the intensity of selected frequency from the spectrum of each pixel. We used the open source data mining software Orange (https://orange.biolab.si) for frequency slicing the hyperspectral data.

**Table 1|** Launcher efficiency for tip and Au edge.

| ω (cm$^{-1}$) | air substrate | | Au substrate | |
| --- | --- | --- | --- | --- |
| | $\sigma_{Tip}$ | $\sigma_{Edge}$ | $\sigma_{Tip}$ | $\sigma_{Edge}$ |
| 1425 | 22 | 0.64 | 0.3 | 2.5×10$^{-3}$ |
| 1433 | 25 | 0.64 | 0.09 | 1.6×10$^{-3}$ |
| 1450 | 4 | 0.04 | 0.16 | 2.0×10$^{-12}$ |
| 1466 | 2.5 | 0.09 | 0.08 | 2.6×10$^{-12}$ |

## Acknowledgements

The authors thank the LNLS for providing beamtime to this project, and LNNano for assistance in the device construction, and Lab Nanomateriais at UFMG for allowing use of the atomic layer transfer system. We thank Yves Petroff for the in-depth discussions and for stimulating the research. Thiago M. Santos, Vinícius O. da Silva and Neaspec GmbH are acknowledged for the technical assistance in the experiments. A.R.C, I. D. B and L. C. C thank the financial supported from CAPES, Fapemig, CNPq and INCT/Nanomateriais de Carbono. C. D. acknowledges financial support by CNPq. A.B. acknowledges financial support through the AFOSR grant FA9550-14-1-0376.

Supplementary Information

# Observation of Diode Behavior and Gate Voltage Control of Hybrid Plasmon-Phonon Polaritons in Graphene-Hexagonal Boron Nitride Heterostructures

Francisco C. B. Maia,*,[1] Brian T. O'Callahan,[2] Alisson R. Cadore,[3] Ingrid D. Barcelos,[1,3] Leonardo C. Campos,[3] Kenji Watanabe,[4] Takashi Taniguchi,[4] Christoph Deneke,[5,6] Alexey Belyanin[7], Markus B. Raschke,[2] Raul O. Freitas*,[1]

[1]Brazilian Synchrotron Light Laboratory (LNLS), Brazilian Center for Research in Energy and Materials (CNPEM), Zip Code 13083-970, Campinas, Sao Paulo, Brazil.

[2]Department of Physics, Department of Chemistry, and Joint Institute for Lab Astrophysics (JILA), University of Colorado, Boulder, Colorado 80309, United States.

[3]Department of Physics, Federal University of Minas Gerais, 30123-970 – Belo Horizonte, Minas Gerais, Brazil.

[4]Advanced Materials Laboratory, National Institute for Materials Science, 305-0044 – Namiki, Tsukuba, Japan.

[5] Brazilian Nanotechnology National Laboratory (LNNano), Brazilian Center for Research in Energy and Materials (CNPEM), Zip Code 13083-970, Campinas, Sao Paulo, Brazil

[6] Applied Physics Department, Gleb Wataghin Physics Institute, University of Campinas (Unicamp), Zip Code 13083-859, Campinas, SP, Brasil.

[7]Department of Physics & Astronomy, Texas A&M University, College Station, Texas 77843-4242, United States.

*Corresponding authors: francisco.maia@lnls.br / raul.freitas@lnls.br

1. The Fano model

The Fano effect[1] for graphene systems[2,3] has been attributed to the interaction of the G band vibration and broadband electronic response giving rise to asymmetric resonances in the infrared absorption spectrum. The analytic expression of the Fano resonance $I_{Fano}$ is given by eq. S1 as a function of the excitation frequency $\omega$, phonon frequency $\omega_0$, damping $\Gamma$, transition dipole strength of coupled mode p, and dimensionless Fano factor f. As commonly understood, f corresponds to the transition dipole strength between the discrete and continuum states. In our systems, the Fano effect arises from the interaction of the hyperbolic phonon-polariton (HPhP) modes of the hBN, which are discrete states, with the surface plasmons from graphene that can be described as broadband electronic oscillations in mid-IR range. Thus, f gives information on the coupling strength of the hybrid hyperbolic plasmon phonon-polariton (HP³) modes.

Here, we extend the Fano resonance (eq. S1), which is given for one oscillator, to multiple discrete resonances (eq. S2). This extended expression $I_{HP^3}$ assumes that the intensity $I_{HP^3}$ results from the squared modulus of the sum over j discrete phonon resonances, and a complex non-resonant background $A_{nr}$[4,5]. Disregarding the background term $A_{nr}$, eq. S2 converges to the eq. S1 as expected.

Particularly, the Fano model assumes that each phonon mode can individually interact with the broadband plasmons, thus each HP³ mode contains an intrinsic Fano factor. Different Fano factors define particular line-shapes for the resonance, as can be seen in Fig. S1. This implies that the inherent interference between distinct adjacent resonances determine the overall intensity.

$$I_{Fano} \propto \frac{p(f.\Gamma + (\omega-\omega_0))^2}{(\omega-\omega_0)^2 + \Gamma^2} \quad \text{eq S1.}$$

$$I_{HP^3} = \left| \sum_j \frac{\sqrt{p_j}\left(f_j \Gamma_j + (\omega-\omega_0)\right)}{(\omega-\omega_0) + i\Gamma_j} + A_{nr} \right|^2 \quad \text{eq S2.}$$

Using the eq. S2, we have fitted, with good agreement, the experimental spectra of G-hBN on distinct substrates and the spectra of G-hBN/Au upon different gate voltages as shown in the paper. The fit procedure consisted in keeping the same number of modes and their respective resonance frequencies for all spectra, while no constraints were put on the other parameters. These criteria reasonably assumed that the substrate influence and the bias neither create novel resonant modes nor cause appreciable shift in the resonant frequency since those properties are mostly associated with the crystalline structure that remains unaltered in our experiments.

Depending on the material substrate, f can assume different values. In particular for the mode at 815 cm$^{-1}$, we find f = -30.5 for G-hBN/Au, f = 9.2 for G-hBN/SiO$_2$ and f = 4.4 for G-hBN/air. The origin of the difference among those fs is addressed in the main paper according to the HP$^3$ coupling on each substrate.

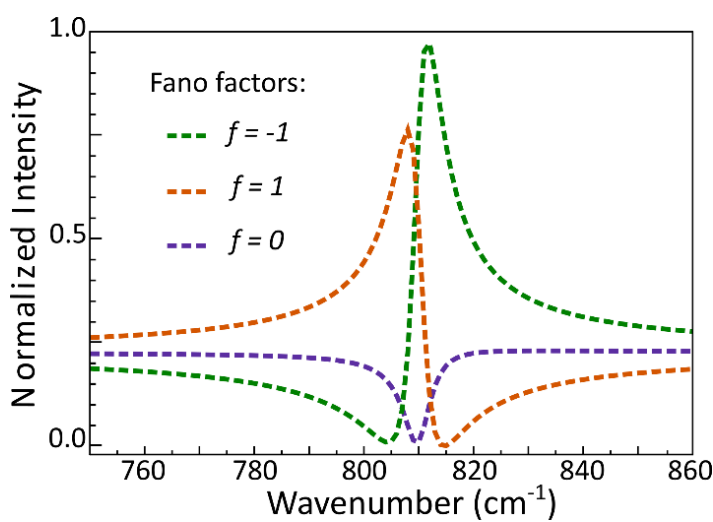

**Figure S1|** Fano-like spectral intensities (eq. 2) are plotted considering three different values of $f$ with $\omega_0$ = 800 cm$^{-1}$, p = 1, $\Gamma$ = 5 cm$^{-1}$ and A$_{nr}$ = e$^{i\pi/4}$. One sees that despite that the parameters associated with the vibrational resonance are kept unaltered, small variations of $f$ leads to very distinguished line-shapes.

## 2. Gate Modulation of the Diode Effect

Spectral linescans on the G-hBN heterostructure bridging Au and air for +4V and -4V gate voltages, respectively, are shown in Fig. S2 a,b. We observe very different shape of the spatio-spectral regions corresponding to the reflection of the HP$^3$s for +4 V (Fig. S2a) and -4V (Fig. S2b). According to our interpretation, such difference signifies a change in the transmitted/reflected HP$^3$ power flow. Moreover, it is noted that gate-induced wavelength selection: the highest amplitude modes for -4 V appear near 780 cm$^{-1}$, whilst, for + 4V the 815 cm$^{-1}$ mode is the strongest.

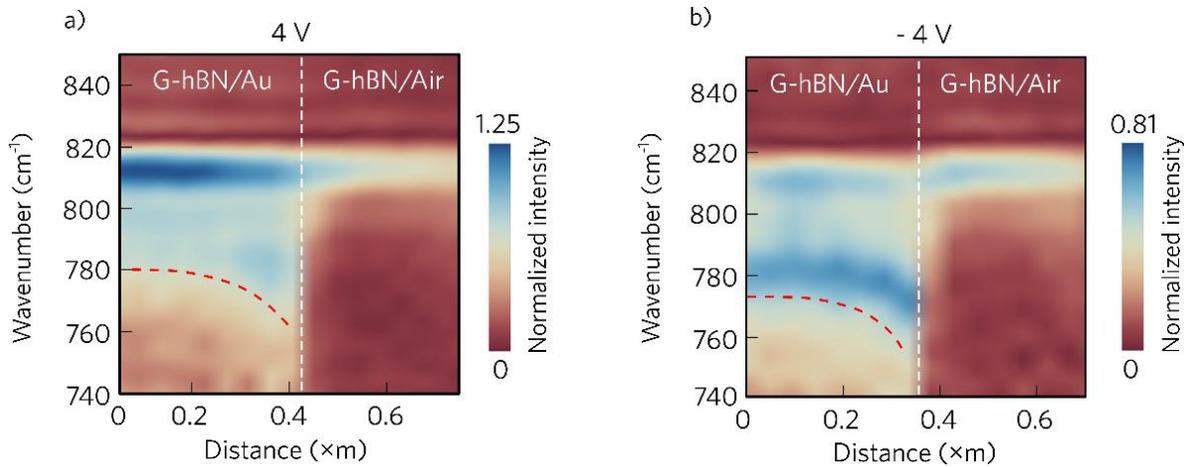

**Figure S2|** Linescan on the G-hBN heterostructure on the Au (left) and Air (right) substrates using 4 V (**a**) and -4 V (**b**) bias.

## 3. Calculated Dispersion Relations and Reflection Coefficient of Type I Band on Au and Air Substrates

Fig. S3a presents the calculated dispersion relation of type I $HP^3$ modes on air and Au substrate.

The calculated reflection coefficient for the type I band (Fig. S3b) indicates the asymmetric reflection coefficient for out-of-plane modes launched on air and Au sides. The understandings on power flow and wave sources for the $HP^3$ modes addressed for the type II band, in the main manuscript, hold true for the type I band by considering an inverted relation for the polariton momentum with respect to the type II band. For type I band, $HP^3$ momenta on air are higher than on Au. Hence, in analogy to type II discussion, there is a momenta mismatch for type I $HP^3$s at the air-Au substrate transition leading to reflection of those modes as shown by the intensity maps and oscillations in the intensity profiles shown, respectively, in Fig. S3c and d.

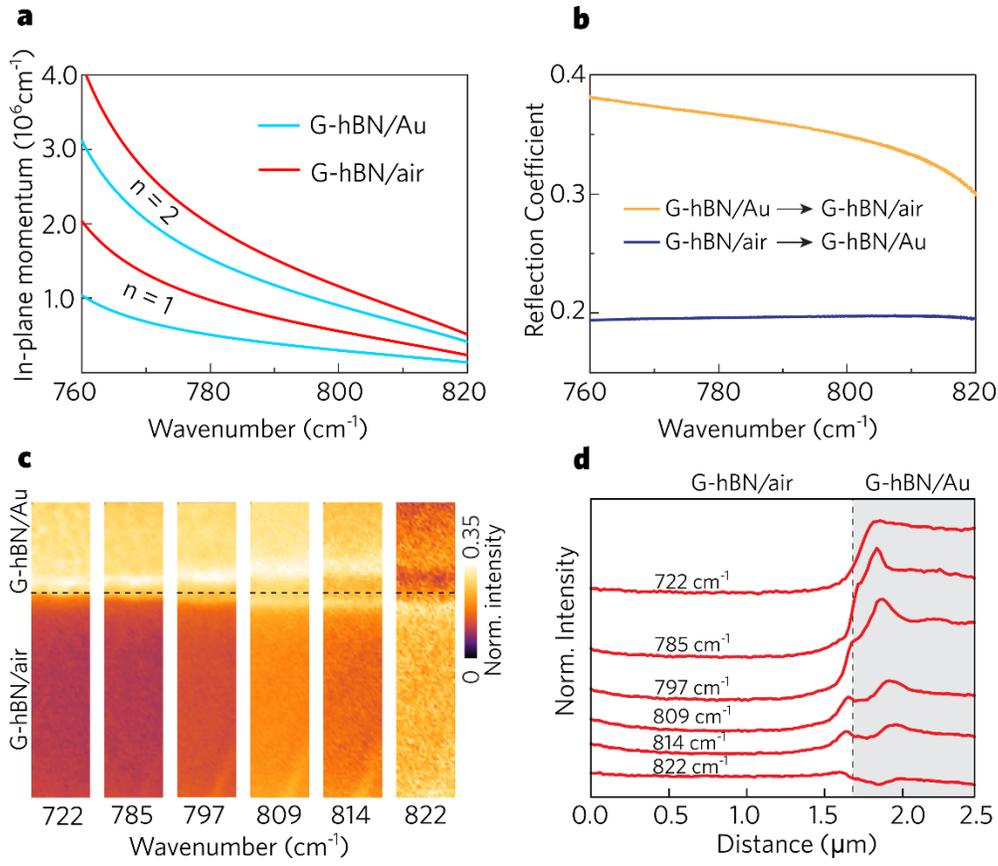

**Figure S3 |** (**a**) Dispersion of the HP$^3$ modes calculated for type I band regarding the two heterostructures. (**b**) Reflection coefficient calculated from eq. 3 for type I HP3 modes propagating from G-hBN/air to the G-hBN/Au heterostructure (blue curve) and in the opposite direction (yellow curve). (**c**) Single-frequency intensity maps of a 35 nm thick G-hBN, on a 0.5 µm x 2.5 µm region with the underneath air-Au metasurface (air-Au transition is marked by the dashed lines), for different excitation wavenumbers in the type I band. (**d**) Intensity profiles extracted from the intensity maps in **d**.

## 4. Hybridization of Au Surface Plasmons and hBN Hyperbolic Phonon-Polaritons

Hybridization of Au surface plasmons (SP) and HPhP of the hBN is shown by the spectral linescan of a hBN crystal lying on Au substrate (Fig. S3a). In the hBN/Au side, the characteristic type I band of hBN/Au is seen, yet, on the Au side a spatio-spectral resonance rises extending from 3.2 µm (crystal edge) to 7.2 µm and with maximum centered near 814 cm$^{-1}$. Comparing with the spectral linescans of hBN/Si (Fig. S3b) and hBN/SiO$_2$ (Fig. S3c), we note that such resonant feature does not appear neither in Si nor SiO$_2$ surfaces. Since the Au substrate is spectrally non-resonant, we attribute such resonance to the SP-HPhP hybridized modes. In this case, on the Au side the tip launches IR broadband SPs that propagate up to the crystal edge wherein form hybrid SP-HPhP modes. By transmission, SP-HPhP modes travel inside hBN/Au. Reflected SP-HPhP modes form the spatio-spectral resonance on the Au side of the Fig. S3a.

These results allow stating the participation of the SP-HPhP hybridization in the HP³ coupling observed in the G-hBN/Au heterostructure.

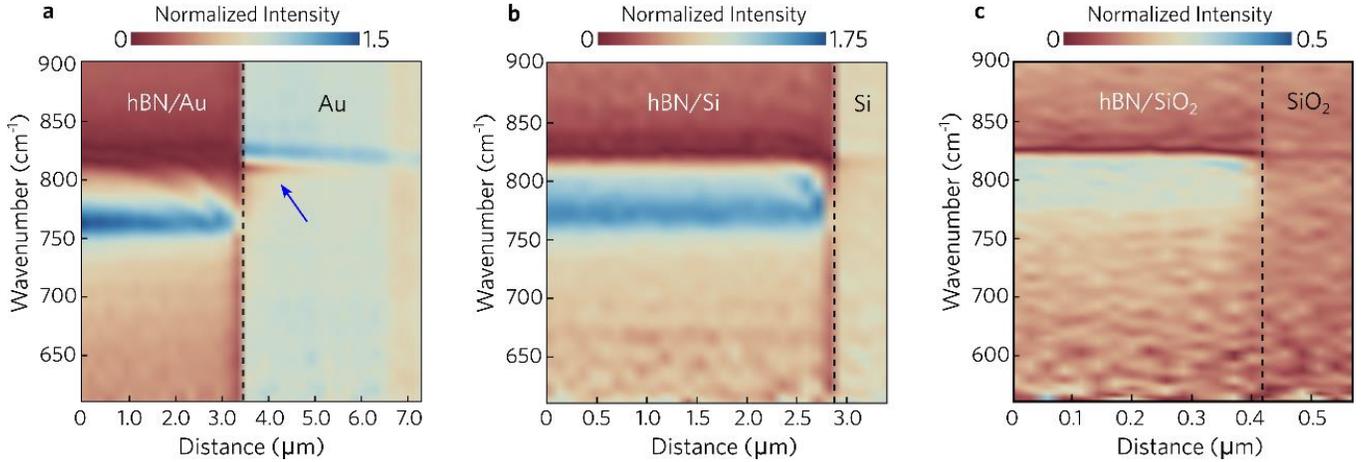

**Figure S4|** Spectral linescans of the type I of hBN crystal lying on Au (**a**), Si (**b**) and SiO$_2$ (**c**) substrates. The vertical dashed lines mark the edge of the crystals. The blue arrow in (**a**) indicates a spatio-spectral resonance, spanning from the crystal edge to 7.2 µm with maximum near 814 cm$^{-1}$.

## 5. Modelling the HP³ waves in the G-hBN Heterostructure Bridging the Au-air Metasurface

The G-hBN/Au and G-hBN/air junction can be seen as a general case of a two-dimension crystal lying onto a metasurface composed of media 2 and 3 (Figure S4). These media influence the dielectric environment creating media 1 and 2 in the crystal. Far-field excitation, typically, illuminates a large area on the sample inducing the antenna effect to all sharped structures into the radiation volume[6]. These antennas confine free-space radiation in the form of near-field and are able to launch polaritonic waves whose type, e.g., circular or plane waves, is determined to the geometry of the structure. The resulting optical field $\xi_{opt}$ is given by the interference of all those waves in the polaritonic medium. The $\xi_{opt}$ mode structure will be defined by the most efficient launcher structures.

In our case, media 1, 2, 3 and 4 are, respectively, G-hBN/air, G-hBN/Au, air and Au. The main launchers are the metalized AFM tip and Au edges. Considering the reference frame in the Figure S4b for the waves in the G-hBN/air, the tip optical field $\xi_{Tip}$ (eq. S3, the same as eq. 4 of the main paper) is given by circular waves $a$ and $a'$ (first and second terms in eq. S3), from the tip at x, that propagate to +/- x directions and reflect, at the junctions at x = 0, L, as $r \times a$ and $r \times a'$ waves (third and fourth terms in eq. S3) carrying the air→Au reflection coefficient $r$. Tip-launched waves have amplitude $A$ and phase $\alpha$, whilst, reflected waves present amplitude $r \times A$ and phase $\alpha + \pi$. The optical field $\xi_{Au}$ (eq. S4, the same as eq. 5 of the main paper) is composed plane waves c and c' (first and second terms in the eq. S4), from the edges at x = 0, L, carrying the amplitude $B$ and phase $\beta$. All waves possess the same complex momentum $Q = q + i\kappa$. Hence, tip and Au edges produce $\xi_{Opt} = \xi_{Tip} + \xi_{Au\ edges}$ which is the near-field scattered by the tip. $|\xi_{Opt}|^2$ is the intensity at the detection whose equation is the used in the fittings of the Fig. 4b of the main paper. Note that this picture remains valid for the tip on the G-hBN/Au with the Au→air reflection coefficient in this case.

$$\xi_{Tip} = A \frac{e^{-i\alpha}}{\sqrt{x}} \left( e^{i(q+i\kappa)x} + e^{-i(q-i\kappa)x} - r \times e^{-i(q-i\kappa)2x} - r \times e^{i(q+i\kappa)2(x_{edge}-x)} \right) \quad \text{eq. S3}$$

$$\xi_{Au} = B e^{-i\beta} \left( e^{-i(q-i\kappa)x} - e^{i(q+i\kappa)2(x_{edge}-x)} \right) + C e^{-i\eta} \quad \text{eq. S4}$$

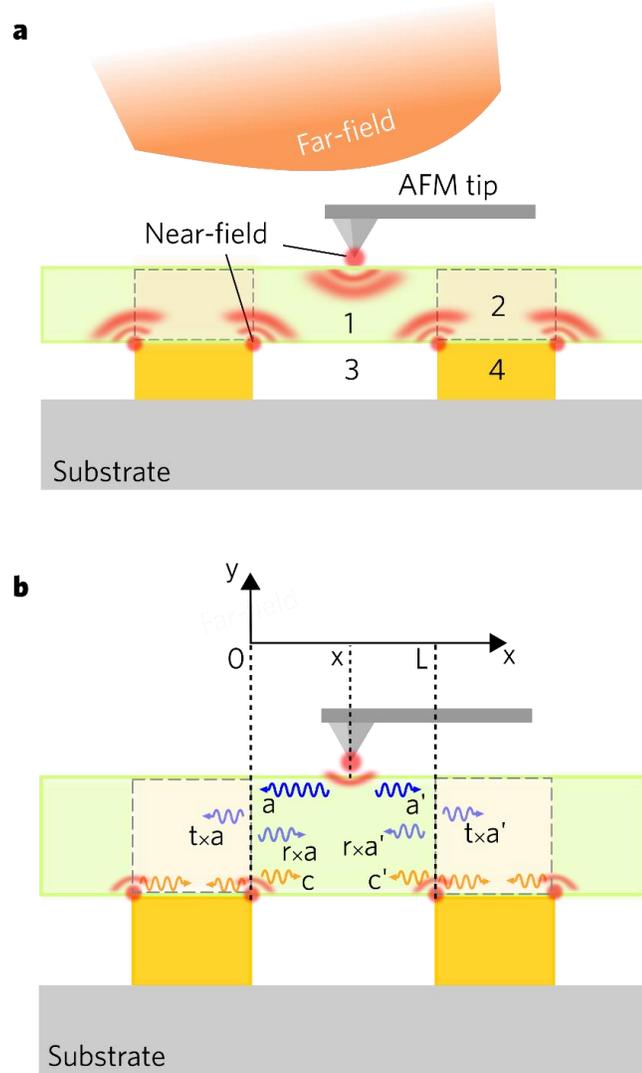

**Figure S5|** (**a**) Scheme of the main antennas responsible for the near-field sources is our system. (**b**) reference frame and representation of the waves taken into account for the $\xi_{Opt}$ modelling. Tip launched circular **a** and **a'** waves propagate in the G-hBN/air to the junction with G-hBN/Au, reflect as circular $r \times$ **a** and $r \times$ **a'** waves at the heterointerface on the substrate transition. The $t \times$ **a** and $t \times$ **a'** waves the transmitted ones to the G-hBN/Au from the incident **a** and **a'**. The waves $c$ and c' are emitted from the Au edges.

The model-extracted amplitudes, complex momenta and reflection coefficients allow us to obtain the launcher efficiency for the tip $\sigma_{Tip} = 2\pi[2A(1+r)]^2$ and the Au edges $\sigma_{Edge} = B^2$, which are addressed in the main manuscript. Likewise, we are able to provide the damping of each type II HP$^3$ wave for the two substrates in the Table S1.

**Table 1|** Damping of the type II HP$^3$ modes for Au and air substrates.

| $\omega$ (cm$^{-1}$) | $\gamma_{air}$ | $\gamma_{Au}$ |
|---|---|---|
| **1425** | 0.02 | $0.1 \times 10^{-3}$ |
| **1433** | 0.03 | $0.4 \times 10^{-3}$ |
| **1450** | 0.003 | $1.0 \times 10^{-3}$ |
| **1466** | 0.011 | $3 \times 10^{-3}$ |